\documentclass[%
reprint, superscriptaddress,
nofootinbib, 
amsmath,amssymb, aps, pra,
]{revtex4-1}%

\usepackage{txfonts}
\usepackage{graphicx}
\usepackage{bm}
\usepackage{hyperref}

\usepackage[english]{babel}
\graphicspath{ {fig/} }
\usepackage{bbold}
\usepackage{braket}
\usepackage[usenames,dvipsnames,svgnames,table]{xcolor}
\usepackage{gensymb}
\usepackage{soul}

\usepackage[capitalize]{cleveref}

\newcommand{\dd}{\mathrm{d}}

\newcommand{\Tr}[1]{\mathrm{Tr}\left\{#1\right\}}

\begin{document}
	
	
	\title{Braiding Majorana corner modes in a second-order  topological 
	superconductor}

	\author{Tudor E. Pahomi}
	\email{pahomit@phys.ethz.ch}
	\affiliation{Institute for Theoretical Physics, ETH Zurich, 8093 Zurich, Switzerland}

	\author{Manfred Sigrist}
	\affiliation{Institute for Theoretical Physics, ETH Zurich, 8093 Zurich, Switzerland}

	\author{Alexey A. Soluyanov}%
	\altaffiliation{Deceased, October 2019}
	\affiliation{	Physik-Institut, University of Zurich, 8057 Zurich, Switzerland}
	\affiliation{	Department of Physics, St. Petersburg State University, St. Petersburg 199034, Russia	}
	
	\begin{abstract}
	We propose the concept of a device based on a square-shaped sample of a two-dimensional second-order topological helical superconductor which hosts two zero-dimensional Majorana quasiparticles at the corners. %
	The two zero-energy modes rely on particle-hole symmetry (PHS) and their spacial position can be shifted by rotating an in-plane magnetic field and tuning proximity-induced spin-singlet pairing. %
	We consider an adiabatic cycle performed on the degenerate ground-state manifold and show that it realizes the braiding of the two modes whereby they accumulate a non-trivial statistical phase $\pi$ within one cycle.
	Alongside with the PHS-ensured operator algebra, the fractional statistics confirms the Majorana nature of the zero-energy excitations.
	A schematic design for a possible experimental implementation of such  a device is presented, which could be a step towards realizing non-Abelian braiding.

	\end{abstract}

	\maketitle


Standard $d$-dimensional topological insulators and superconductors have a gapped bulk spectrum and exhibit conducting surface states in $(d-1)$ dimensions~\cite{Kane2005Z2,Bernevig2006,Fu2006,Fu2007,Fu2007a,Hasan2010,Qi2011,Chiu2016}. The analysis of non-interacting electrons in materials possessing fundamental symmetries -- time-reversal ($\mathcal{T}$), particle-hole ($\mathcal{P}$) and/or chiral symmetry ($\mathcal{C}$) -- led to the initial classifications of such 
topological phases~\cite{Schnyder2008,Schnyder2009,Kitaev2009}.
Later on, possible topological phases protected by discrete crystalline symmetries were studied~\cite{Fu2011,Hsieh2012,Slager2012,Chiu2013,Morimoto2013,Shiozaki2014,Benalcazar2014,Alexandradinata2014a,Shiozaki2016,Kruthoff2017,Shiozaki2017,Thorngren2018,Khalaf2018a} and, lastly, another class of exotic non-interacting topological phases was discovered, namely the second-order topological insulators (SOTIs) and superconductors (SOTSs)~\cite{Benalcazar2017,Song2017,Schindler2018,Geier2018,Khalaf2018,Ahn2018,Miert2018}.

Unlike standard gapped topological materials (that may be termed "first-order"), such systems have insulating $(d-1)$-dimensional surfaces, but host topologically-protected boundary modes in $(d-2)$ dimensions. In particular, for $d=3$ this means the material exhibits one-dimensional (1D) states propagating along its \emph{hinges}
\cite{Benalcazar2017b,Song2017,Langbehn2017,Miert2018,Schindler2018,Wang2018a,Calugaru2019,Ghorashi2019,Agarwala2020} 
as detected in Bi(111) \cite{Schindler2018a}.
Accordingly, SOTIs and SOTSs 
in $d=2$ host localized ($d=0$) \emph{corner modes} and have been discussed in
both static
\cite{Benalcazar2014,Benalcazar2017,Benalcazar2017b,Langbehn2017,Wang2018a,Miert2018,Calugaru2019,Volpez2019,Agarwala2020}
	and Floquet-driven systems \cite{Franca2018,Rodriguez-Vega2019,Huang2020}.
Second-order topological phases have been experimentally realized in artificial systems, such as  mechanical metamaterials \cite{Serra-Garcia2018}, microwave \cite{Peterson2018} and topolectrical circuits \cite{Imhof2018}, and further implementations in materials have been proposed  \cite{Schindler2018,Ezawa2018,Wang2018a,Hsu2018}.

One of the motivations to seek for topological materials is their potential implementation in devices that would exploit the topology-ensured protection from decoherence. %
Of particular interest is \textit{topological quantum computation}, considered the cornerstone of quantum technology, 
in which logical operations are done by means of braiding (double exchange) of non-Abelian quasiparticle excitations~\cite{Kitaev2001,Kitaev2006,Nayak2008,Alicea2012,Sarma2015}. Majorana (zero-energy) excitations  are known to be the only non-Abelian quasiparticles that can appear in the absence of electron-electron interactions~\cite{Fidkowski2010,Fidkowski2011a} 
and were initially predicted to appear at the edges of a 1D 
spinless superconductor~\cite{Kitaev2001}. Almost a decade later, it was demonstrated that this exotic topological phase could be achieved in a semiconductor nanowire with strong spin-orbit coupling placed in an external magnetic field, if superconductivity is induced by proximity effect
~\cite{Lutchyn2010,Oreg2010}. Following these theoretical proposals, such devices were experimentally realized by several groups~\cite{Mourik2012, Das2012, Deng2012, Finck2013, Churchill2013, Deng2016} and
are currently among the most promising building block for topological quantum computing devices.  %
However, despite the existing proposals of braiding Majorana end modes using 
nanowires~\cite{Alicea2011, Heck2012}, 
identifying the topological qubit remains extremely challenging due to the engineering complexity of manipulating the topological phase in the wire junctions without losing the 
superconducting state. This motivates the search for other approaches to realize a Majorana qubit in potentially feasible condensed matter systems~\cite{Zhu2018,Yan2018,Wang2018,Liu2018a,Schrade2018}.

\begin{figure}[t]
	\centering
	\includegraphics[width=\linewidth]{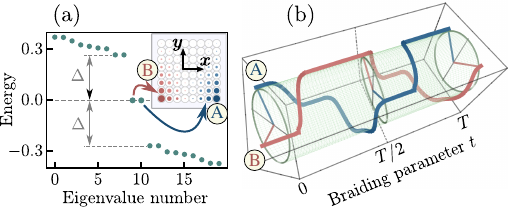}
	\caption{
		(a) Energy of the lowest 20 eigenstates in a finite square-shaped sample of our second-order topological superconductor for parameters  $\bm b =(-0.3,0)$ and $\bm s = (0,0.3)$; the two zero-energy modes, denoted by A and B, are separated from the bulk states by an energy gap $\Delta$ and the probability densities $|\psi_\alpha(x,y)|^2$  are schematically shown in the inset ($\alpha\in\{\mathrm{A,B}\}$). %
		(b) Worldlines of A and B during the proposed cyclic adiabatic process of period $T$, computed as 	$\arg\max_\varphi|\psi_\alpha(\varphi)|^2$ for each $t\in[0,T]$, with $\varphi$ the polar angle.}	
	\label{latt_cyl}
\end{figure}

\newcommand{\ABenpsi}{Fig.~\ref{latt_cyl}a}
\newcommand{\ABcyl}{Fig.~\ref{latt_cyl}b}
\newcommand{\AB}{Fig.~\ref{latt_cyl}}


In this paper we propose a device based on a two-dimensional (2D) SOTS which hosts two Majorana modes localized at two corners of a square-shaped crystalline sample (see \ABenpsi). We call them Majorana corner states (MCSs) in the following. 
Within our model, the corner-localization of the MCSs in the device can be tuned using an in-plane magnetic field and proximity-induced spin-singlet superconductivity. In this setup, the braiding of the Majorana quasiparticles can be realized in a straightforward fashion (schematically illustrated in \ABcyl), and we show that the associated fractional statistics of the proposed topological Majorana qubit is observed.





\newcommand{\grdiff}{-}

\section*{SOTS model}

\label{Model}

In $\bm k$-space, our proposed model can be formulated with a minimal four-band Bogoliubov-de Gennes Hamiltonian 
\begin{equation}
	\mathcal{H}(\bm k) = \begin{pmatrix} 
		\hat{h}_{\bm k} & \hat{\Delta}_{\bm k} \\
		\hat{\Delta}^*_{\bm k} & - \hat{h}^*_{\bm k}
	\end{pmatrix},
	\label{eq:H}
\end{equation}
written in the Nambu basis $\hat{\Psi}_{\bm k} :=\left(\begin{matrix}c_{\bm{k}\uparrow}&c_{\bm{k}\downarrow}&c_{-\bm{k}\uparrow}^\dagger&- c_{-\bm{k}\downarrow}^\dagger\end{matrix}\right)^T$, with $c^{(\dagger)}_{\bm k\,s}$ being the annihilation (creation) operator for an electron with a 2D momentum $\bm k$ and spin $s\in \{\uparrow,\downarrow\}$. The gap function considered is (with $\bm k$-labels suppressed)
\begin{equation}
	\begin{split}
		\hat{\Delta}
		=
		\left(\begin{matrix}\Delta_{\uparrow\uparrow} & -\Delta_{\uparrow\downarrow}\\\Delta_{\downarrow\uparrow} & -\Delta_{\downarrow\downarrow}\end{matrix}\right),
		%
		\quad
		\begin{array}{l}
			{\Delta}_{\uparrow\uparrow} = t_0(\Delta_{\downarrow\downarrow}^* =\! \sin k_x - i \sin k_y)\, ,\\
			\Delta_{\downarrow\uparrow} = \!-\Delta_{\uparrow\downarrow}=\! s_x \cos k_x+s_y \cos k_y\,,
		\end{array}
	\end{split}
	\label{Helems}
\end{equation}
while the diagonal blocks are given by
\begin{equation}
	\hat{h}_{\bm k} =  t_0(1-\cos{k_x} - \cos{k_y})\sigma_0 + \bm b \cdot \boldsymbol{\sigma}\,,
\end{equation}
where $\bm\sigma=(\sigma_1,\sigma_2,\sigma_3)$ and  $\sigma_0$ are the three Pauli matrices and identity  in spin space, respectively. 
The vectors $\bm b \equiv (b_x,b_y)$ and $\bm s \equiv (s_x,s_y)$ parametrize the in-plane magnetic field and the spin-singlet pairing amplitudes, respectively, with magnitudes smaller than $t_0$, taken as unit in the following.


\begin{figure}[t]
	\centering
	\includegraphics[width=\linewidth]{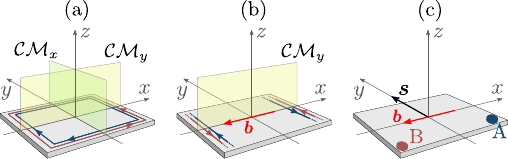}
	\caption{%
		(a) Schematic diagram for the helical state with two counter-propagating edge modes  of opposite spins. At the $x$ ($y$) boundary,  they  are protected by the mirror antisymmetry $\mathcal{CM}_y$ ($\mathcal{CM}_x$), a product of chiral symmetry and the mirror reflection $\mathcal{M}_y$ ($\mathcal{M}_x$).
		(b) The magnetic field $\bm b=(-0.3,0)$ breaks the mirror antisymmetry $\mathcal{CM}_x$ and gaps the $y$ edges. 
		(c) If, additionally, the parameter $\bm s =(0,s_y)\parallel \hat{\bm y}$ is non-vanishing, 	the pairing term $\Delta_{\bm k,\downarrow\uparrow}=s_y \cos k_y$ breaks the $\mathcal{CM}_y$ antisymmetry and two zero-energy modes appear, localized at two adjacent corners.
	}		
	\label{mirrors}
\end{figure}

\newcommand{\mirrxy}{Fig.~\ref{mirrors}a}
\newcommand{\mirry}{Fig.~\ref{mirrors}b}
\newcommand{\mirr}{Fig.~\ref{mirrors}c}


For $\bm b = \bm s = \bm 0$, the Hamiltonian $\mathcal{H}_{(\bm 0,\bm 0)}$
describes electrons with nearest-neighbor hopping on a square lattice, subject
to $p$-wave pairing (for clarity, we use here the vectors $\bm b$ and $\bm s $ as
	subscripts to refer to a specific parameter configuration).
 $\mathcal{H}_{(\bm 0,\bm 0)}$  possesses all three non-crystalline symmetries ($\mathcal{T}$, $\mathcal{P}$ and $\mathcal{C}$) and falls into the symmetry class DIII~\cite{Schnyder2008,Schnyder2009,Kitaev2009}, supporting helical Majorana edge states (\mirrxy). Including crystalline symmetries, the total point symmetry group is 
 \begin{equation}
 	\mathcal{G}_{(\bm 0,\bm 0)} = D_{4h} \times \{ \mathbb{1},
 	\mathcal{T}, \mathcal{P}, \mathcal{C} \}\,,
 \end{equation} 
where $\mathbb{1}$ represents the identity. The main symmetries elements of
	$\mathcal{G}_{(\bm 0,\bm 0)}$ are listed in \cref{tablesymm}, alongside with
	the corresponding representations. Since $\mathcal{P}$ is always present in
	our model, $\mathcal{G}_{(\bm 0,\bm 0)}$  can be rewritten as:%
\begin{equation}
	\mathcal{G}_{(\bm 0,\bm 0)} =\tilde{\mathcal{G}}_{(\bm 0,\bm 0)}\times \{
	\mathbb{1}, \mathcal{P} \}\,,
\label{G00total}
\end{equation}
where the magnetic group for $\bm b=\bm s=0$ reads $\tilde{\mathcal{G}}_{(\bm
	0,\bm 0)}= D_{4h} \oplus \mathcal{T} D_{4h}$ (or $4/mmm1'$). The magnetic group is reduced for other values of the parameters $\bm b$ and $\bm s$, as explained below, and detailed in \cref{app:symmetries}.

A finite in-plane magnetic field $\bm b$ conserves $\mathcal{P}$ for any direction, but breaks $\mathcal{T}$. The reduced magnetic group is $\tilde{\mathcal{G}}_{(\bm b,\bm 0)}= C_{2h}\oplus \mathcal{T}(D_{2h}\grdiff C_{2h})$ \footnote{Here, the set $(D_{2h}\grdiff C_{2h})$ represents the relative complement of $C_{2h}$ with respect to $D_{2h}$. It contains the elements of $D_{2h}$ which are not part of $C_{2h}$ and it is \emph{not} a group by itself.}  if $\bm b$ is parallel to $\hat{\bm x}$, $\hat{\bm y}$ or $\hat{\bm x}\pm\hat{\bm y}$, and $C_i\oplus\mathcal{T}(C_{2h}\grdiff C_i)$  otherwise (i.e. $m'm'm$ and $2'/m'$, respectively). The magnetic field hybridizes the helical boundary modes unless they are protected by symmetry: as illustrated  in \mirry, for $\bm{b}  \parallel \hat{\bm x}$  only the $x$ edges remain metallic protected by $\mathcal{CM}_y \in \mathcal{G}_{(\hat{\bm x}, \bm 0)}$, a product of $\mathcal C$ and the mirror reflection $\mathcal{M}_y$ which maps $(x,y,z)\mapsto(x,-y,z)$. Analogously, for $\bm {b} \parallel \hat{\bm y}$ the gapless states at the $y$ edges are protected by $\mathcal{CM}_x$ 
(see \cref{app:protection} for a proof). 
Both edges are gapped for generic orientations $\bm{b}\nparallel \hat{\bm x},\hat{\bm y}$, but, if $\bm{b}\parallel(\hat{\bm x} \pm\hat{\bm y})$, the mirror-symmetric corners bisected by $\bm b$ will host zero-energy modes~\cite{Geier2018,Ezawa2018b}.

The spin-singlet pairing controlled by $\bm s$ ($|\bm s| \neq 0$) reduces the symmetry further and in the generic case $\tilde{\mathcal{G}}_{(\bm b,\bm s)}=C_1\oplus\mathcal{T}(C_s\grdiff C_1)$ (or $m'$). Thus, the (minimal) total group of the total Hamiltonian is $\mathcal{G}_{(\bm b,\bm s)} = \{\mathbb{1},\mathcal{TM}_z\}\times \{ \mathbb{1}, \mathcal{P} \}$, including two particularly important elements, $\mathcal{P}$ and the effective chiral symmetry $\widetilde{\mathcal{C}}=\mathcal{CM}_z=\mathcal{PTM}_z$, which transform the Hamiltonian in the following way:
\begin{subequations}
	\begin{align}
	\mathcal H(\bm k)&= -\; U_\mathcal{P} \;\mathcal H(-\bm k)^*U_\mathcal{P}^\dagger  &U_\mathcal{P}& = \sigma_3\tau_1
	\label{PHS}\\
	\mathcal H(\bm k) &= -\; U_{\widetilde{\mathcal{C}}} \;\mathcal H(+\bm k) \;U_{\widetilde{\mathcal{C}}}^\dagger  &U_{\widetilde{\mathcal{C}}}& = \sigma_2\tau_1
	\label{eCS}
	\end{align}
\end{subequations}
where the Pauli matrix $\tau_1$  acts on the particle-hole space.

\section*{Majorana corner states} 
\label{MCS}

On a finite square-shaped lattice, the $\mathcal{P}$-symmetric Hamiltonian hosts two degenerate zero-energy excitations $\ket{\psi_\alpha}$, with $\alpha \in \{\mathrm{A},\mathrm{B}\}$, whose features in  the second-order topological phase are displayed in \ABenpsi~for $\bm b = (-0.3,0)$ and $\bm s = (0,0.3)$. In this regime,  although the magnetic field $\bm b \parallel \hat{\bm x}$ would allow for $\mathcal{CM}_y$-protected boundary modes at the $x$ edges, the spin-singlet pairing controlled by $\bm s \parallel \hat{\bm y}$ gaps the $x$ edges by breaking this symmetry (\mirr) and enforces the zero-energy modes to localize at two adjacent corners of the device (as shown in the inset of \ABenpsi). More precisely, the pairing term $\Delta_{\bm k,\downarrow\uparrow}$ 
reduces	$\tilde{\mathcal{G}}_{(\hat{\bm x},\bm 0)}=C_{2h}\oplus \mathcal{T}(D_{2h}\grdiff C_{2h})$ to $\tilde{\mathcal{G}}_{(\hat{\bm x},\hat{\bm y})} = C_s \oplus\mathcal{T}(C_{2v}\grdiff C_s)$ (i.e. $m'm'm$ to $m'm2'$).

The operators corresponding to the MCSs can be written as
\begin{equation}
	\gamma_\alpha =  \sum_{\bm r}\sum_{i=1}^4 \left[\psi_{\alpha,\bm r}\right]^*_i \left[ \hat{\Psi}_{\bm r}\right]_i\,,\quad \alpha \in \big\{\text{A},\text{B}\big\}\,,
	\label{gamma}
\end{equation}
where $[\hat{\Psi}_{\bm r}]_i$ is the $i$-th component of the Fourier-transformed Nambu basis for the site at position $\bm r$ and $\left[\psi_{\alpha,\bm r}\right]_i$ the $i$-th component of the corresponding sector $\ket{\psi_{\alpha,\bm r}}$ of the MCS $\ket{\psi_\alpha}$. Because $\ket{\psi_\alpha}$ are self-conjugate under $\mathcal P$ ($\mathcal{P}\ket{\psi_{\alpha,\bm r}} \equiv U_\mathcal{P}\ket{\psi_{\alpha,\bm r}}^* =  \ket{\psi_{\alpha,\bm r}}$), these operators satisfy the following relations (for details see \cref{app:realspace})
\begin{equation}
\gamma_\alpha = \gamma_\alpha^\dagger\,,\qquad\{\gamma_\alpha,\gamma_\beta\} = 2\delta_{\alpha\beta}\,.
\label{Malgebra}
\end{equation}

The energies of the midgap states go to zero exponentially with the system size 
(data shown in \cref{app:realspace}), are robust against $\mathcal{P}$-preserving disorder (tested numerically, data not shown) and are well separated from the bulk states by an energy gap (emphasized in \ABenpsi). The localization of the  boundary modes is strongly dependent on the Hamiltonian parameters $\bm b$ and $\bm s$, whereby their energy remains zero for all the considered configurations, in which the device is in a SOTS phase. This degenerate ground state manifold is crucial to realize the braiding of the topological MCSs.


\begin{figure}[t]
	\centering
	\includegraphics[width=\linewidth]{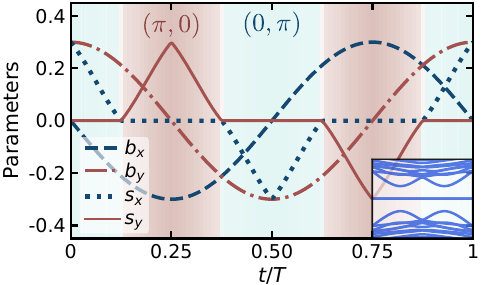}
	\caption{The evolution of the Hamiltonian parameters $\bm b \equiv (b_x,b_y)$ and $\bm s \equiv (s_x,s_y)$ during the cyclic adiabatic process of period $T$. The alternating dark-light background colors correspond to second-order topological phases with  invariants $(\phi_{xy},\phi_{yx})=(\pi,0)$ or $(0,\pi)$, respectively.
		The inset displays the energy eigenvalues of the finite lattice in the range $E\in[-0.4,0.4]$,	during the last quarter of the cyclic process (the period of the eigenvalues is $T/4$).
	}
	\label{fig:bsinvar}
\end{figure}

\newcommand{\bsinvar}{Fig.~\ref{fig:bsinvar}}

\section*{Computing topological invariants}
\label{TopInv}

We can assess the topological features of our device using the 
Wilson loop operator (WLO). Several technical aspects are explained in \cref{app:WLO}.

The WLO can be adiabatically connected to the Hamiltonian at the boundary~\cite{Blount1962,King-Smith1993,Fidkowski2011,Taherinejad2014} and can thus characterize the surface topology of materials, such as $\mathbb{Z}_2$ topological (crystalline or not) insulators in 2D and 3D~\cite{Fu2006,Soluyanov2011,Soluyanov2011a,Yu2011,Taherinejad2014,Alexandradinata2014,Alexandradinata2016,Gresch2017} or $\mathbb{Z}$ Chern insulators~\cite{Coh2009,Qi2011a,Alexandradinata2014,Gresch2017}.

Moreover, the \emph{nested} WLO introduced in Ref.~\cite{Benalcazar2017} has been used to address the physics at "the boundary of the boundary"~\cite{Schindler2018,Franca2018,Xie2018}. In our 2D model, using the nested WLOs we obtain two  $\mathbb{Z}_2$ corner topological invariants $\phi_{xy},\phi_{yx} \in \{0,\pi\}$, which are quantized by the effective chiral symmetry introduced in Eq.~\eqref{eCS} (confirmed by numerical simulations). The device is in a SOTS phase whenever either of $\{\phi_{xy},\phi_{yx}\}$ is equal to $\pi$; for example, in the case $\bm b=(-0.3,0)$ and $\bm s =(0,0.3)$, shown in \ABenpsi, we have $(\phi_{xy},\phi_{yx}) = (\pi,0)$.


\begin{figure}[t]
	\centering
	\includegraphics[width=\linewidth]{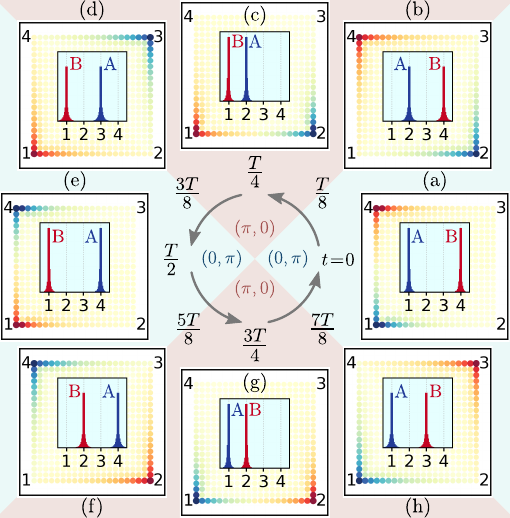}
	\caption{Square lattices with $21\times21$ sites, displaying the spatial probability density $P_\text{A}(t;x,y)$ (blue color) and  $P_\text{B}(t;x,y)$ (red color) of the two zero-modes A and B during the adiabatic cycle, at time $t_\text{a}=0$ in panel (a), $t_\text{b}=T/8$ in panel (b), $t_\text{c}=2T/8$ in panel (c) etc. %
	The probability of localization is maximum at the corners of the lattice (labeled with 1, 2, 3 and 4), which can be seen from the intensity of the colors. This feature is emphasized in the insets, where  we used the parametrization $(x,y) = r(\cos\varphi,\sin\varphi)$ and plotted the angular  probability distributions $P_\alpha(t;\varphi)$ for $\alpha \in \{\text{A},\text{B}\}$.
	The background colors are alternating when $t \in [0,T]$ and have the same meaning as in \bsinvar: the second-order topological invariants $(\phi_{xy},\phi_{yx})$ are equal to $(0,\pi)$ in  the light regions and to $(\pi,0)$ in the darker regions.
	}	
	\label{fig:snapshots1}	
\end{figure}

\newcommand{\psiT}{Fig.~\ref{fig:snapshots1}}


\section*{Braiding process}
\label{Braiding}

We perform the adiabatic braiding of the MCSs by rotating the magnetic 
field ($\bm b$) and adjusting the spin-singlet pairing ($\bm s$).  Let $T$ be the time period of the \textit{cyclic} process and $t \in [0,T]$ the  parameter which controls the process, such that we have periodic boundary conditions $\mathcal H^{(t=0)}=\mathcal H^{(t=T)}$. The evolution of the parameters $\bm b(t)$ and $\bm s(t)$ over a full period is shown in \bsinvar, where the alternations of the second-order topological states characterized by $(\phi_{xy},\phi_{yx}) = (\pi,0)$ and $(0,\pi)$ are also highlighted. Along this path in parameter space, the bulk and the edges remain gapped, while the energy of the degenerate MCSs is unaffected -- they  remain well-separated from the bulk states by an energy gap (feature displayed in the inset of \bsinvar). The MCSs thus do not mix with the bulk states during the cyclic process, which we illustrate in \psiT~by computing the spatial probability density of $\ket{\,\psi_\alpha(t)\,}$  in the proposed device 
for several $t$ ($\alpha\in\{\text{A},\text{B}\}$).


\section*{Statistical phase}
\label{StatPhase}

A fermionic many-body wavefunction picks a factor $(-1)$ upon exchanging 
two fermions and it reverts to the initial wavefunction if the exchange 
of any other two is performed. In contradistinction, in a system 
with $2M$ Majorana quasiparticles ($M>1$) two subsequent exchanges 
(say, $\gamma_1\leftrightarrow \gamma_2$ and $\gamma_2\leftrightarrow \gamma_3$) 
generally give a different state (non-Abelian braid group) 
\cite{Kitaev2006,Nayak2008,Alicea2012,Sarma2015}. 
In particular, if the system contains a single pair of Majoranas ($M=1$), 
such a double exchange (braiding) reveals the \emph{fractional statistics} 
of the two quasiparticles: each acquires a non-trivial statistical phase $\Phi = \pi$. 

The device we propose here realizes the braiding of two midgap states
$\psi_\alpha$ [with corresponding operators $\gamma_\alpha$, $\alpha
\in\{\text{A},\text{B}\}$,
see Eq.~\eqref{gamma}] and we show below their statistics 
is indeed that of Majorana quasiparticles (fractional, $M=1$). 
We note, however, that in the proposed setup  the single exchange 
$\text{A}\leftrightarrow \text{B}$ 
is not feasible with the parameter cycle considered in \bsinvar.
Although at $t=T/2$ the centers of the wavefunctions 
$\psi_\text{A}$, $\psi_\text{B}$ are swapped  (see \psiT),
the Hamiltonian after a half-period is different from the initial one.
Consequently, a statistical phase between 
$\psi_\text{A}$ ($\psi_\text{B}$) at $t=0$ and
$\psi_\text{B}$ ($\psi_\text{A}$) at $t=T/2$ 
would not be well-defined nor robust.

To express the braiding phase accumulated by each MCS within one full cycle
(i.e. after a double exchange), 
we use the Berry phase defined as
\begin{equation}
\Phi_\alpha = \oint_0^{T}\left[\mathcal{A}_{\;t}\right]_{\alpha\alpha}\, \mathrm{d}t\,,\quad \alpha \in \{\text{A, B}\}\,,
\label{eq:StatPhase}
\end{equation}
where the Berry connection matrix has elements $\left[\mathcal{A}_{\;t}\right]_{\alpha\beta} = i\bra{\,\psi_\alpha(t)\,} \partial_t \ket{\,\psi_\beta(t)\,}$ 
(more details are given in \cref{app:statphase}).
The quantity in Eq.~\eqref{eq:StatPhase} is gauge-invariant because  
(i) the Hamiltonian parameter configurations at  $t\in\{0,T\}$ are identical and (ii)
the matrix $\mathcal{A}_{\;t}$ is diagonal $\forall t \in [0,T]$, as 
the two MCSs are kept far apart from each other and do not mix during 
the course of the adiabatic cycle. 
The practically null spatial overlap of the two states A and B 
was checked numerically and is illustrated in \ABcyl~and \psiT.

With Eq.~\eqref{eq:StatPhase}, we obtain non-trivial statistical phases $\Phi_\text{A}=\Phi_\text{B}=\pi$, which means
\begin{equation}
\left\{\begin{array}{c}
{\gamma_\text{A}} \quad\to\quad - \; {\gamma_\text{A}} \\
{\gamma_\text{B}} \quad\to\quad - \; {\gamma_\text{B}}
\end{array}\right.,
\label{eq:psi12braid}
\end{equation}
and, together with the operator algebra~\eqref{Malgebra}, 
proves the Majorana nature of the quasiparticles. 
As confirmed numerically,
the quantization of the phases is guaranteed by $\mathcal{P}$, Eq.~\eqref{PHS}.

However, the relative phase described here would be an inaccessible 
experimental quantity within the proposed protocol, since the 
two MCSs necessarily have different fermion parities.
Nonetheless, we believe it might be possible to extend our scheme as to
construct a system hosting $M>1$ Majorana pairs (for example, by 
distributing $M$ devices as a chain or a two-dimensional array).
In such a setup, 
we would expect that the fractional statistics~\eqref{eq:psi12braid} 
of our MCSs ($M=1$)
would generalize to a  non-Abelian 
braid group.

\begin{figure}[t]
	\centering
	\includegraphics[width=\linewidth]{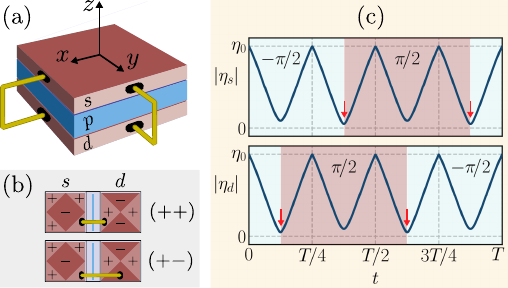}
	\caption{(a) Schematic representation of the proposed device, consisting of a two-dimensional helical $p$-wave superconductor in contact with an extended $s$-wave and a $d$-wave superconductor. %
	(b) Diagrams of the spin-singlet gap structures in the $xy$ plane, with positive and negative sectors. One of the Josephson contacts between the $s$ and $d$ superconductors (depicted as well in panel a) links two sectors of the same sign (++ configuration), while the other links sectors of different signs (+-). %
	 (c) The magnitudes of the spin-singlet gap functions during the proposed adiabatic cycle of period $T$, with $+\pi/2$ ($-\pi/2$) phases relative to the $p$-wave superconductor  in the $t$-intervals with dark-red (light-blue) background color. One commutes  between the (++) and (+-) configurations when both $|\eta_{s/d}|$ are small; the red arrows indicate the smaller gap magnitude and thus the spin-singlet superconductor which flips the phase relative to the $p$-wave. %
	}
	\label{fig:device}	
\end{figure}

\newcommand{\device}{Fig.~\ref{fig:device}a}
\newcommand{\sdlobes}{Fig.~\ref{fig:device}b}
\newcommand{\etasd}{Fig.~\ref{fig:device}c}


\section*{Schematic device design} %

\label{DevDesign}

Finally, we briefly describe a possible structure 
for the proposed device that would consist of
three square-shaped thin-film superconductors stacked as in \device. %
The superconducting state in our model ($\hat{\Delta}_{\bm k}$)
can be realized if we induce
spin-singlet pairing in the $p$-wave film through proximity effect, 
by placing it between an extended $s$-wave superconductor and a 
$d_{x^2-y^2}$-wave superconductor, 
with even-parity gap functions 
$\psi_{s/d}=\eta_{s/d}(\cos k_x \pm \cos k_y)$. 
In terms of the $\bm{s}$%
-parameter, the gap magnitudes can be written as 
$\eta_{s/d} = (s_x \pm s_y)/2$. %
The relative phases between the superconductors can be fixed by Josephson contacts, as explained in the following.  %

First, the order parameter $\eta_p$ corresponding to the $p$-wave couples only in second order to the spin-singlet ones, which locks the $p$-$s$ and $p$-$d$ relative phases to either $\pi/2$ or $-\pi/2$ (in order to minimize the Josephson energy, in lowest order being typically $E_\text{J} \propto + \cos 2\phi_{p-s/d}$).  Indeed, first-order coupling terms as $(\eta_{s/d}^*\,\eta_p + \text{c.c.})$ are assumed to be absent or extremely weak, due to the mismatch of parity. Note that the $p$-wave superconducting state, $d$-vector $\bm d_{\bm k} = \hat{\bm x} \sin k_y- \hat{\bm y} \sin k_x$, belongs to irreducible representation $A_{2u}$, while the  order parameters $\psi_s$ and $\psi_d$ belong to the even-parity representations $A_{1g}$ and $B_{1g}$, respectively. Thus, the phase coupling between the odd-parity and even-parity order parameter would be dominated by the second order coupling of the form $(\eta_{s/d}^{*2}\,\eta_p^2 + \text{c.c.})$. %
The appropriate phase factor $e^{i\pi/2}$ is already considered in Eq.~\eqref{eq:H}, where the configuration "$\eta_s + \eta_d + i \eta_p$" is presented.

Second, the relative phase between the spin-singlet condensates can be controlled using alternatively two direct Josephson contacts (see \device).
The gap functions $\psi_s$ and $\psi_d$ exhibit in the $xy$ plane positive and negative sectors, 
as schematically represented in \sdlobes, and
the Josephson contacts are set up such that they link equal- or different-sign sectors, respectively. 
Switching the active contact increases by $\pi$ not only the relative $s$-$d$ phase, but consequently also \emph{that} singlet-triplet phase, where the Josephson coupling is weaker. %
As emphasized in \etasd, during the proposed adiabatic cycle such an operation changes the $p$-$s$ ($p$-$d$) phase by $\pi$ when $|\eta_s|<(>)|\eta_d|$.

Manipulating the spin-singlet gap magnitudes $\eta_{s/d}$ during the proposed cycle remains a challenging task. At a conceptual level, two strategies could be taken under consideration: using tunable tunneling Josephson contacts between the $p$-wave and the spin-singlet superconductors or modifying the order parameters amplitudes, $ \eta_s $ and $ \eta_d $ individually by local heating or other means. In fact, small changes to the proposed functions $\eta_{s/d}(t)$, that might come within the experimental implementation, are not expected to change the topological features of the cycle. The latter argument applies as well to the in-plane magnetic field $\bm{b}(t)$. Moreover, we assume this to be weak enough 
such that in the superconducting thin films
the interference with the Josephson effect as well as the depairing effects are negligible.


\vspace{1em}

\textit{Note. The physical system considered
in the \href{https://arxiv.org/abs/1904.07822v2}{previous arXiv version (1904.07822v2)} differs from the current one, 
however, 
their topological and symmetry features are closely related.
}

\begin{acknowledgments}
	
T.E.P. would like to thank J.L. Lado, T. Kawakami and K. Viebahn for helpful discussions. T.E.P. and M.S. are grateful for the financial support from the Swiss National Science Foundation (SNSF) through Division II (No. 163186 and 184739). A.A.S. acknowledges the support of Microsoft Research, SNSF NCCR MARVEL and QSIT programs, and the SNSF Professorship grant.
	
\end{acknowledgments}



\begin{appendix}

\section{Symmetries of the model Hamiltonian}

\label{app:symmetries}

The group $\tilde{\mathcal{G}}_{(\bm	0,\bm 0)}$ of 
\cref{G00total} is a Heesch-Schubnikov magnetic group of type II; a magnetic field ($|\bm b|\neq 0$) breaks $\mathcal{T}$ and reduces it 
to a magnetic group of type III. 
	In the general case (possibly $|\bm s|\neq0$), we can write
\begin{equation}
	\tilde{\mathcal{G}}_{(\bm b, \bm s)}
	= {N}_{(\bm b, \bm s)} \oplus \mathcal{T} \left(G_{(\bm b, \bm s)} -{N}_{(\bm b, \bm s)}\right)
	\,, 
	\label{GHSH}
\end{equation}
where ${G}_{(\bm b, \bm s)}$ is a subgroup of $D_{2h}$ and ${N}_{(\bm b, \bm s)}$ a halving (and normal) subgroup of ${G}_{(\bm b, \bm s)}$. The following relations between the subgroups of $D_{2h}$ hold:
\begin{equation}
	D_{2h} \;(mmm) \supset \left\{
	\begin{array}{lll}
		C_{2v} \;(mm2) & \supset C_{1h} \;(m) 		& \supset C_1 \;(1)	\\
		C_{2h} \;(2/m) & \supset C_{1h} \;(m) 		& \supset C_1 \;(1)	\\
		C_{2h} \;(2/m) & \supset C_i \;(\bar{1})	& \supset C_1 \;(1)	
	\end{array}
	\right.
\end{equation}%
The groups ${G}_{(\bm b, \bm s)}$ and ${N}_{(\bm b, \bm s)}$ are given 
	explicitly in \cref{tablesymm} for certain values of $\bm b$ and $\bm s$; 
	in those six cases, the magnetic point groups  
	$\tilde{\mathcal{G}}_{(\bm b, \bm s)}$ in international notation are $m'm'm$,\;\, $m'm2'$,\;\, $2'/m'$,\;\, $m'$,\;\, $m'm'm$,\;\, $m'm2'$. 
		In the Table, we denote by $\{\tau_j\}$, with $j\in\{0,1,2,3\}$, the set of Pauli matrices  acting in particle-hole space.


\begin{figure}[t]
	\centering
	\includegraphics[width=\linewidth]{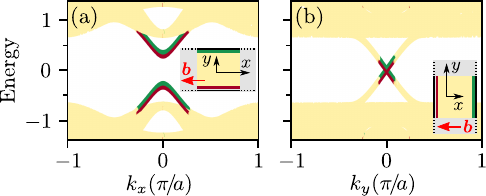}
	\caption{%
		Edge spectra in the presence of the magnetic field $\bm b \parallel \hat{\bm x}$. Panel (a) illustrates the spectrum of the $y$ boundaries, while panel (b) describes the $x$ edges. The color of each eigenvalue corresponds to the localization of the respective eigenstate: the boundary modes (green and red) can be clearly distinguished from the bulk states (light yellow).}		
	\label{ribbons}
\end{figure}

\newcommand{\helical}{\mirrxy}
\newcommand{\mirrxapp}{\mirrxy}
\newcommand{\mirryapp}{\mirry}
\newcommand{\ribbBx}{\cref{ribbons}a-b}


\newcommand{\cmark}{\checkmark}%
\newcommand{\xmark}{$\times$}
\newcommand{\plotwidth}{44pt}
\newcommand{\sizeBs}{\small}

\setlength\extrarowheight{3.5pt}
\setlength{\tabcolsep}{5pt}

\begin{table*}[t]\centering\begin{tabular*}{\textwidth}{l|llr|cc|cc|cc}\hline\hline
		&&
		&\!\!\!\includegraphics[width=39pt]{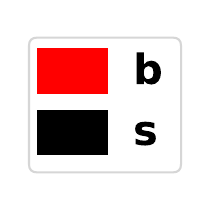} &
		\includegraphics[width=\plotwidth]{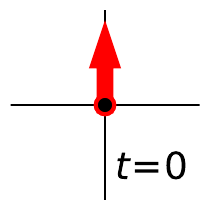}&\includegraphics[width=\plotwidth]{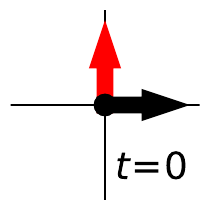}  & \includegraphics[width=\plotwidth]{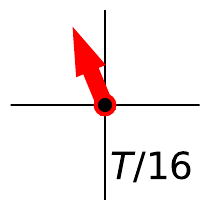}&\includegraphics[width=\plotwidth]{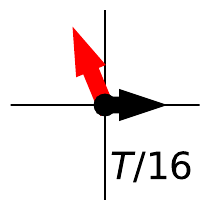}  &   \includegraphics[width=\plotwidth]{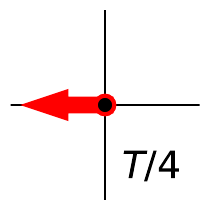}&\includegraphics[width=\plotwidth]{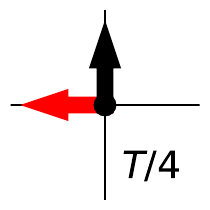}\\

		&Equation of $\mathcal{S}$&&
		& {\sizeBs $\bm{b}\!=\!(0,1)$ }&{\sizeBs $\bm{s}\!=\!(1,0)$} &{\sizeBs $\bm{b}\!=\!(-0.5,1)$ }& {\sizeBs $\bm{s}\!=\!(0.5,0)$ }&{\sizeBs $\bm{b}\!=\!(-1,0)$ }& {\sizeBs $\bm{s}\!=\!(0,1)$ } 	\\
		
		$\mathcal{S}$&$\mathcal{H}(k_x,k_y)\,\;\reflectbox{\rotatebox[origin=c]{90}{$\Lsh$}}$  &$U\,\;\reflectbox{\rotatebox[origin=c]{90}{$\Lsh$}}$&
		
		$\!\!{G},{N}$ {\footnotesize$\to$} & $D_{2h},C_{2h}$ & $C_{2v},C_{1h}$ & $C_{2h},C_i$ &  $C_{1h},C_1$ &      $D_{2h},C_{2h}$ & $C_{2v},C_{1h}$ \\		
		
		\hline

		$\mathcal{P}$ & $-U\mathcal{H}(-k_x,-k_y)^*U^\dagger \; $ &$\sigma_3\tau_1$	&  & \cmark & \cmark &  \cmark & \cmark &  \cmark & \cmark\\ 
		
		$\mathcal{T}$ & $+U\mathcal{H}(-k_x,-k_y)^*U^\dagger$ &$i\sigma_2\tau_0$		&  & \xmark & \xmark &  \xmark & \xmark &  \xmark & \xmark\\
		
		$\mathcal{C}$ & $-U\mathcal{H}(k_x,k_y)U^\dagger$ &$\sigma_1\tau_1$			&   & \xmark & \xmark &  \xmark & \xmark &   \xmark & \xmark\\

		\hline
		$\mathcal{I}$   & $+U\mathcal{H}(-k_x,-k_y)U^\dagger$ &$\sigma_0\tau_3$		&  & \cmark & \xmark &  \cmark & \xmark &  \cmark & \xmark\\
		
		$\mathcal{R}_{2}^x$ & $+U\mathcal{H}(k_x,-k_y)U^\dagger$&$\sigma_1\tau_3$	&  & \xmark & \xmark &  \xmark & \xmark &  \cmark & \xmark\\
		
		$\mathcal{R}_{2}^y$ & $+U\mathcal{H}(-k_x,k_y)U^\dagger$ &$i\sigma_2\tau_0$	&  & \cmark & \xmark &  \xmark & \xmark &   \xmark & \xmark\\
		
		$\mathcal{R}_{2}^z$ & $+U\mathcal{H}(-k_x,-k_y)U^\dagger$ &$\sigma_3\tau_3$	&  & \xmark & \xmark &  \xmark & \xmark &  \xmark & \xmark\\
		

		$\mathcal{M}_{x}$  & $+U\mathcal{H}(-k_x,k_y)U^\dagger$ &$\sigma_1\tau_0$	&  & \xmark & \xmark &  \xmark & \xmark &  \cmark & \cmark\\
		
		$\mathcal{M}_{y}$ & $+U\mathcal{H}(k_x,-k_y)U^\dagger$ &$i\sigma_2\tau_3$	&  & \cmark & \cmark &  \xmark & \xmark &   \xmark & \xmark\\
		
		$\mathcal{M}_z$ & $+U\mathcal{H}(k_x,k_y)U^\dagger$ &$\sigma_3\tau_0$		&  & \xmark & \xmark &  \xmark & \xmark &  \xmark & \xmark\\

		\hline
		
		$\mathcal{T}\,\cdot$		 &  &		&  &  &  &   &  &   & \\

		$\cdot\;\mathcal{R}_2^x$ & 
		&$\sigma_3\tau_3$& &\cmark & \cmark & \xmark & \xmark &\xmark& \xmark\\
		
		$\cdot\;\mathcal{R}_2^y$ &
		&$\sigma_0\tau_0$&&\xmark & \xmark &  \xmark &\xmark & \cmark & \cmark\\

		$\cdot\;\mathcal{R}_2^z$ & 
		&$\sigma_1\tau_3$&& \cmark & \xmark & \cmark & \xmark & \cmark & \xmark\\

		$\cdot\;\mathcal{M}_{x}$ &
		&$\sigma_3\tau_0$&&\cmark & \xmark & \xmark & \xmark & \xmark & \xmark\\

		$\cdot\;\mathcal{M}_{y}$ & 
		&$\sigma_0\tau_3$& &\xmark & \xmark &  \xmark & \xmark &\cmark&\xmark\\

		$\cdot\;\mathcal{M}_z$ & 
		&$\sigma_1\tau_0$&&  \cmark & \cmark &  \cmark & \cmark  & \cmark & \cmark\\
		
		\hline

		$\mathcal{C}\,\cdot$		&  &		&  &  &  &   &  &   & \\

		$\cdot\;\mathcal{M}_{x}$ & 
		&$\sigma_0\tau_1$&  & \cmark & \xmark &  \xmark & \xmark &   \xmark &\xmark\\
		
		$\cdot\;\mathcal{M}_{y}$ & 
		&$i\sigma_3\tau_2$&  & \xmark & \xmark &  \xmark & \xmark &   \cmark & \xmark\\
		
		$\cdot\;\mathcal{M}_z$ & 
		&$i\sigma_2\tau_1$&  & \cmark & \cmark &  \cmark & \cmark &   \cmark & \cmark\\

		\hline\hline
		
	\end{tabular*}
	
	\caption{	Symmetries ($\mathcal{S}$) of the bare Hamiltonian $\mathcal{H}_{(\bm 0,\bm 0)}$, on lines: particle-hole antiunitary antisymmetry ($\mathcal{P}$), time-reversal antiunitary symmetry ($\mathcal{T}$), chiral antisymmetry ($\mathcal{C}=\mathcal{PT}$),  inversion symmetry ($\mathcal{I}$), two-fold rotation (around $j$ axis, $\mathcal{R}_{2}^j$), mirror reflection ($\mathcal{M}_j$ maps $x_j\mapsto-x_j$), antiunitary symmetries ($\mathcal{TR}_2^j$ and $\mathcal{TM}_j$) and unitary antisymmetries ($\mathcal{CM}_j$).
		For each symmetry, the equation of transformation is given, as well as the corresponding representation $U$ in Hamiltonian space.
		In the following six columns it is indicated whether the Hamiltonian $\mathcal{H}_{(\bm b,\bm s)}$ still possesses a certain symmetry, for various values of $\bm{b}=(b_x,b_y)$ and $\bm{s}=(s_x,s_y)$ -- depicted as arrows of different colors in the table header.
		The parameters for three representative steps of the braiding process
		are considered, $t \in\{0,T/16,T/4\}$. For each step, two cases are compared:
		[$\bm{b}$ as indicated and $\bm{s}=(0,0)$] in the former column 
		and [$\bm{b}$ as indicated, $\bm{s}$ as indicated] in the latter column. 
		For each column, the groups  ${G}$ and ${N}$ explained in	\cref{GHSH} are specified.
	}
	\label{tablesymm}
\end{table*}


\section{Edge states and their symmetry protection}

\label{app:protection}

The two counter-propagating edge modes of opposite spins 
hosted by the system described by $\mathcal{H}_{(\bm{0},\bm{0})}$
(helical state, depicted in \helical) continue to exist 
on certain edges even when the magnetic field is switched on, 
while they aquire a gap on other edges. 
We prove their symmetry protection by introducing a one-dimensional (1D) winding number.

Consider a 2D system with a finite number of unit cells in $k_\perp$ direction, but with periodic boundary conditions in the other direction $k_\parallel$. Provided the bulk Hamiltonian $\mathcal{H}(\bm k)\equiv \mathcal{H}(k_x,k_y)$ has an effective chiral symmetry (with representation $\Gamma$) for a certain $k_\parallel$,
\begin{equation}
	\left\{\Gamma, \mathcal H(k_\parallel = k_{0},k_\perp) \right\} =0\,,\qquad \forall k_\perp\,,
	\label{chiralGamma}
\end{equation}
then the topology of the point $k_0$ is characterized by the following 1D winding number \cite{Sato2011}:
\begin{equation}
	W_\text{1D} \left(k_\parallel\right) = - \frac{1}{4\pi i } \oint \dd k_\perp \Tr{\Gamma \mathcal{H}(\bm k)^{-1} \partial_{k_\perp} \mathcal{H}(\bm k)}.
	\label{W1D}
\end{equation}
The chiral symmetry \eqref{chiralGamma} implies that the matrix diagonalizing $\Gamma$ off-diagonalizes the Hamiltonian:
\begin{equation}
	\begin{split}
		U_\Gamma^\dagger \,\Gamma \,U_\Gamma = \mathrm{diag}(-1,-1,1,1) 
		\quad \Rightarrow \quad
		U_\Gamma^\dagger \mathcal H(\bm k) U_\Gamma  = \begin{pmatrix} 0 & f_{\bm{k}} \\ f_{\bm{k}}^\dagger & 0  \end{pmatrix},
	\end{split}
	\label{diagGamma}
\end{equation}
such that \cref{W1D} can be simplified:
\begin{equation}
	W_\text{1D}\left(k_\parallel\right)= \frac{1}{2\pi} \oint \dd k_\perp  \partial_{k_\perp} \arg\left(\det f_{\bm k}\right).
	\label{W1Dh}
\end{equation}

This expresses the number of times the phase of $\det f_{\bm{k}}$ winds when the parameter $k_\perp$ completes a full period.

\textbf{System infinite in the $y$-direction.} We consider the case with magnetic field $\bm{b} = (b,0)\parallel \hat{\bm x}$, for which there are boundary modes only at the $x$ edges, dispersing as $\pm k_y$ at low energies (see \ribbBx). For our model, the chiral symmetry $\mathcal{CM}_z$ is present at any momentum $\bm{k}$:
\begin{equation}
	\begin{split}
		\mathcal{CM}_z:\quad &\left\{\Gamma, \mathcal H(k_x,k_y) \right\} =0 \ ,
		\\
		&\Gamma = \tau_1\sigma_2\,,
		\quad  U_\Gamma=\frac{1}{\sqrt{2}}
		\left(\begin{matrix}0 & i & 0 & - i\\- i & 0 & i & 0\\1 & 0 & 1 & 0\\0 & 1 & 0 & 1\end{matrix}\right).
		\label{GammaCMz}
	\end{split}
\end{equation}

Applying $U_\Gamma$ of \cref{GammaCMz} in \cref{diagGamma} and using ${\Delta}_{\bm k,\uparrow\uparrow} = \Delta_{\bm k,\downarrow\downarrow}^*$, we  obtain the off-diagonal block:
\begin{equation}
	\begin{split}
		f_{\bm{k}}=&\left(
		\begin{array}{cc}
			-\xi_{\bm{k}} & i\,\Big( b  - \Delta_{\bm{k},\uparrow\uparrow}^*\Big) \\
			-i\,\Big(b+\Delta_{\bm{k},\uparrow\uparrow} \Big)& -\xi_{\bm{k}}
		\end{array}
		\right)
		\\&\Rightarrow\quad
		\det f_{\bm{k}} = \xi_{\bm{k}}^2 +|\Delta_{\bm{k},\uparrow\uparrow}|^2 -2i b \sin k_y-b^2\,.
	\end{split}
\end{equation}
According to \cref{W1Dh}, the winding number  at $k_y=0$ is trivial, since $\det f_{(k_x,0)}  = 1-b^2$ is a constant. Therefore, the boundary modes are \emph{not} protected by $\mathcal{CM}_z$. 

We now consider the effective chiral symmetry (see \mirryapp)%
\begin{equation}
	\begin{split}
		\mathcal{CM}_y:\quad& \left\{\Gamma, \mathcal H(k_x,k_y=0) \right\} =0 \ ,
		\\&
		\Gamma = \tau_2\sigma_3\,,
		\quad  U_\Gamma=\frac{1}{\sqrt{2}}
		\left(\begin{matrix}i & 0 & - i & 0\\0 & - i & 0 & i\\1 & 0 & 1 & 0\\0 & 1 & 0 & 1\end{matrix}\right) .
		\label{CMy}
	\end{split}
\end{equation}
This gives a nontrivial winding number as long as $|\bm b|<1$, which means $\mathcal{CM}_y$ protects the boundary modes:
\begin{equation}
	\begin{split}
		\det f_{(k_x,0)} = 
		&\left|
		\begin{array}{cc}
			e^{-i k_x} & b \\
			b & e^{-i k_x} \\
		\end{array}
		\right| =   e^{-2 i  k_x} - b^2
		\\ & \stackrel{\text{\cref{W1Dh}}}{\Longrightarrow}\quad W_\text{1D}(k_y=0)=-2\,.
	\end{split}
\end{equation}
The 1D (edge) system has a $\mathbb{Z}$ topological classification \cite{Schnyder2008,Geier2018} and the topological invariant is precisely the computed winding number, which is equal to the number of states in \emph{one} direction.

\textbf{System infinite in the $x$-direction.} The calculations are similar for perpendicular magnetic field $\bm{b} = (0,b)\parallel \hat{\bm y}$, for which we have edge states at the $y$ edges dispersing as $\pm k_x$ at low energies. Using the symmetry $\mathcal{CM}_z$, at $k_x=0$ we obtain $\det f_{(0,k_y)}= 1- b^2 +2ib\sin k_y$, which gives as well a  trivial winding number. The edge states are protected by $\mathcal{CM}_x$ (see \mirrxapp):
\begin{equation}
	\begin{split}
		\mathcal{CM}_x:\quad &\left\{\Gamma, \mathcal H(k_x=0,k_y) \right\} =0 \ ,
		\\&
		\Gamma = \tau_1\sigma_0\,,
		\quad  U_\Gamma=\frac{1}{\sqrt{2}}
		\left(\begin{matrix}-1 & 0 & 1 & 0\\0 & -1 & 0 & 1\\1 & 0 & 1 & 0\\0 & 1 & 0 & 1\end{matrix}\right) .
	\end{split}
\end{equation}

This gives $\det f_{(0,k_y)} = e^{2 i  k_y} - b^2$ and thus winding number $W_\text{1D}(k_x=0)=2$ if $|\bm b|<1$.


\section{Majorana corner states}

\label{app:realspace}

Consider a squared-shaped sample of a square lattice with $N\times N$ sites. With the Fourier transform ($\bm r$ goes over the positions of all unit cells)
\begin{equation}
	c_{\bm k \sigma} = \frac{1}{N} \sum_{\bm{r}} e^{i\bm{k}\cdot\bm{r}}c_{\bm r\sigma}\,,
\end{equation}
we obtain the superconducting Hamiltonian $\mathcal{H}$ in the basis
\begin{equation}
	\hat{\Psi}_{\bm r} :=\left(\begin{matrix}c_{\bm{r}\uparrow} & c_{\bm{r}\downarrow} & c_{\bm{r}\uparrow}^\dagger &- c_{\bm{r}\downarrow}^\dagger\end{matrix}\right)^T\,.
	\label{Nambu}
\end{equation}
%
%

The Nambu basis $\hat\Psi$ ($\hat\Psi_{\bm k}$ or $\hat\Psi_{\bm r}$) 
is $\mathcal{P}$-symmetric, satisfying
\begin{equation}
	\mathcal{P} \hat\Psi \equiv U_\mathcal{P}\hat\Psi^* =  \hat\Psi,\quad U_\mathcal{P} =\tau_1 \sigma_3\,.
	\label{PHSbasis}
\end{equation}

\begin{figure}[t]
	\centering
	\includegraphics[width=\linewidth]{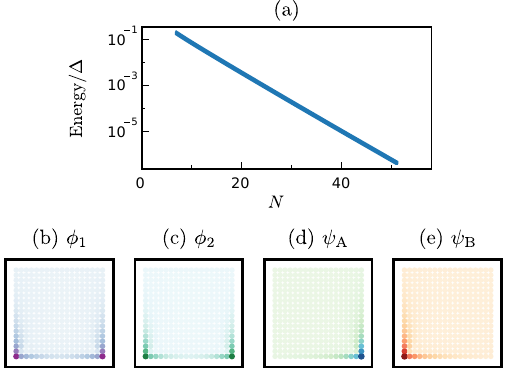}
	\caption{(a) The energy of the two in-gap states, normalized by the bulk energy gap, as a function of the linear system size ($N$). For the case $N=21$, panels (b-c) show the spatial probability densities of the two zero-modes $\ket{\phi_{1,2}}$ and (d-e) of the Majorana corner states $\ket{\psi_{\text{A,B}}}$ (see text). Since the Hamiltonian parameters are chosen at time $t=T/4$ of the cyclic process, by overlapping panels (d-e) we would recover \psiT c.}
	\label{zeromodes}
\end{figure}

\newcommand{\zeroen}{\ABenpsi}

\newcommand{\enN}{\cref{zeromodes}a}
\newcommand{\zerophi}{\cref{zeromodes}b-c}
\newcommand{\zeropsi}{\cref{zeromodes}d-e}

We diagonalize the real-space $(4N^2)\times(4N^2)$ Hamiltonian,
\begin{equation}
	\mathcal{H} \ket{\phi_n}  = E_n \ket{\phi_n}, 
\end{equation}
and plot the spectrum $\{E_n\}$ in \zeroen. The system exhibits two zero-energy modes, which we denote $\ket{\phi_1}$ and $\ket{\phi_2}$. Their energy goes to zero exponentially with increasing $N$ (see \enN),  scaling like $\exp(-N/N_0)$, with $N_0$ a constant. Each of the zero modes is localized on {two} adjacent corners (\zerophi). However, the following two orthogonal linear combinations of the degenerate {fermionic} states $\{\ket{\phi_1},\ket{\phi_2}\}$
\begin{equation}
	\begin{split}
		\ket{\psi_\text{A}} &=\ket{\phi_1} + \ket{\phi_2}\,, \\
		\ket{\psi_\text{B}} &=i\,\big(\ket{\phi_1} - \ket{\phi_2}\big)\,, \\
	\end{split}
\end{equation}
are localized on separate corners (illustrated in \zeropsi). Within our model, it is always possible to build such states $\{\ket{\psi_\alpha}\}$, with $\alpha \in \{\mathrm{A},\mathrm{B}\}$, which satisfy
\begin{equation}
	\mathcal{P}\ket{\psi_\alpha}=\tilde{U}_\mathcal{P}\ket{\psi_\alpha}^* \stackrel{}{=}  \ket{\psi_\alpha}.
	\label{PHSstate}
\end{equation}
The Majorana corner states (MCSs) $\ket{\psi_\alpha}$ are thus self-conjugate under $\mathcal{P}$; together with \cref{PHSbasis}, this property guarantees the operator algebra is that of Majorana excitations \cite{Elliott2015} presented in
\cref{Malgebra}.

\section{Wilson loop operators}

\label{app:WLO}

We used the {nested} Wilson loop operator (WLO) to calculate the second-order topological invariants $\phi_{xy}$ and $\phi_{yx}$. This technique was introduced and explained in detail in Refs.~\cite{Benalcazar2017,Benalcazar2017b}; here we only outline the essential steps.\\

\textbf{The first-order WLO.} First we diagonalize the four-band Bloch Hamiltonian and collect the $N_\text{occ}=2$ occupied states in a matrix:
\begin{equation}
	\mathcal H \ket{u^j_{\bm k}}  = E_{\bm k}^j \ket{u^j_{\bm k}}, \qquad 
	U_{\bm k}=
	\left(\begin{array}{c|c}
		\ket{u^1_{\bm k}} & \ket{u^2_{\bm k}}
	\end{array}
	\right).
\end{equation}
In this way, the projector to occupied space reads
\begin{equation}
	P_{\bm k} := \sum_{i=1}^{N_\text{occ}} \ket{u^i_{\bm k}} \bra{u^i_{\bm k}} = U_{\bm k} U^\dagger_{\bm k}\,.
	\label{projop}
\end{equation}
For the $N\times N$ square lattice with lattice constant $a$, we analyze the adiabatic evolution of a state $\ket{u^m_{\bm k}}$ along a closed path in reciprocal space, with $k_y$ constant, into a final state $\ket{u^n_{\bm k + (2\pi/a,0)}}$. We discretize the path into $N$ points $\bm k_j = \bm k + j \Delta_x$, where 
\begin{equation}
	\Delta_x = (\Delta,0),\qquad \Delta = \frac{2\pi}{Na}.
\end{equation}


\begin{figure}[t]
	\centering
	\includegraphics[width=\linewidth]{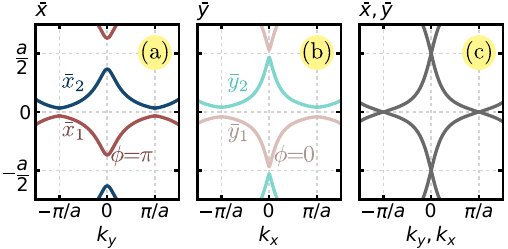}
	\caption{%
		(a) Wannier charge centers $\bar{x}_n$  and (b) $\bar{y}_n$ for the second-order topological superconductor $\mathcal{H}_{(\bm b,\bm s)}$ with parameters $\bm b =(-0.3,0)$ and $\bm s = (0,0.3)$. The Berry phase of the "occupied" Wannier band $\bar{x}_1(k_y)$ in panel (a) is $\phi_{xy}=\pi$, while the $\bar{y}_n(k_x)$ centers  in panel (b) are trivially gapped: $\phi_{yx}=0$. Panel (c) shows the Wannier centers [either $\bar{x}_n(k_y)$ or $\bar{y}_n(k_x)$] for the bare Hamiltonian $\mathcal{H}_{(\bm 0,\bm 0)}$, i.e. parameters $\bm b=\bm s =(0,0)$. In this case, there is no Wannier gap, which reflects the metallic character of all edges in the helical state.
	}		
	\label{WCC}
\end{figure}

\newcommand{\wccxapp}{\cref{WCC}a}
\newcommand{\wccyapp}{\cref{WCC}b}
\newcommand{\wccxyapp}{\cref{WCC}a-b}
\newcommand{\wccapp}{\cref{WCC}c}
\newcommand{\wccsapp}{\cref{WCC}}

If we iteratively project the initial state until the final state is reached, the amplitude of this parallel transport process can be written as:
\begin{equation}
	\left[\mathcal{W}_{x,\bm k}\right]_{nm} := \lim_{N\to\infty}\bra{u^n_{\bm k + (2\pi/a,0)}} 
	{\overline{\prod_{\bm k_i}}} P_{\bm k_i}
	\ket{u^m_{\bm k}},
\end{equation}
where the symbol $\overline\prod$ stands for ordered product. The unitary matrix $\mathcal{W}_{x,\bm k}$ represents the WLO in the $x$-direction; introducing here \cref{projop}, it reads
\begin{equation}
	\begin{split}
		\mathcal{W}_{x,\bm k} =&\lim_{N\to\infty} \overline{\prod_{j}}  O_{x,\bm k + j\Delta_x} =\lim_{N\to\infty} \overline{\prod_{j}} \left( \mathbb{1} +i \Delta\, \mathcal{A}_{\bm k + j\Delta_x}^x \right) \\= &  \,\overline{\exp}\left(i\oint_{q_y =k_y}\mathrm{d} q_x\; \mathcal{A}^x_{\bm q}  \right),
		\label{eq:wlo1app}
	\end{split}
\end{equation}
where $\mathcal{A}^x_{\bm k}$ represents the $x$ component of the Berry connection constructed with the occupied Bloch  eigenfunctions, $\Big(\mathcal{A}^x_{\bm k}\Big)_{nm}=i\bra{u^n_{\bm k}}\partial_{k_x} \ket{u^m_{\bm k}}$. We define the overlap matrices $O_{x,\bm k}$ with elements
\begin{equation}
	\begin{split}
		[O_{x,\bm k}]_{nm} :=& \braket{u^n_{\bm k}|u^m_{\bm k-\Delta_x}} 
		\\ =&
		\bra{u^n_{\bm k}} \Big( \ket{u^m_{\bm k}} -\Delta\, \partial_{k_x} \ket{u^m_{\bm k}}\Big) + \mathcal{O}(\Delta^2) 
		\\ =&\;\delta_{nm} + i\Delta [\mathcal{A}^x_{\bm k}]_{nm}+ \mathcal{O}(\Delta^2).
		\label{Oxk}
	\end{split}
\end{equation}
The last equality of \cref{eq:wlo1app} represents the formal definition of the WLO, although in practice the first equality therein is used (with a finite number $N$ of $\bm k$ points). The operator $\mathcal{W}_{x,\bm k}$ characterizes the $x$ boundary~\cite{Coh2009,Yu2011,Qi2011a} and, for this reason, we may refer to it as "the edge Hamiltonian"  \footnote{The Hermitian edge Hamiltonian $H_{\text{ edge}}$ could be related to the \emph{unitary} WLO through the mapping $\mathcal{W}_{x,\bm k} = \exp\left(i H_{\text{ edge}}\right)$.}. This becomes clear if one proceeds to obtaining  the spectrum of the WLO by solving
\begin{equation}
	\mathcal{W}_{x,\bm k}\ket{\bar{x}_n(\bm k)} = \lambda_{x}^n(k_y)\ket{\bar{x}_n(\bm k)}.
\end{equation}

The eigenvalues $\lambda_{x}^n$ are related to the surface charge polarization \cite{Blount1962,King-Smith1993}, since their phases (non-Abelian Berry phases, $\phi_x^n$) are proportional to the Wannier charge centers (WCCs) $\bar{x}_n$:
\begin{equation}
	\lambda_{x}^n(k_y) =: \exp\left(i \phi_x^n(k_y)\right) = \exp\left(2\pi i\frac{\bar{x}_n(k_y)}{a} \right).
\end{equation}
Here, the quantity $\bar{x}_n$ is defined $\mod\,a$, as the center of the $n$-th hybrid Wannier function, which is maximally localized in $x$-direction and Bloch-like in $k_y$-direction \cite{Soluyanov2011}. Typical shapes for the flow of WCCs $\bar{x}_n$ in our model are displayed in \wccsapp~($n$ goes from $1$ to $N_\text{occ}=2$).

\textbf{Symmetry protection of the WCC crossings.} Let us assume first  $\bm{s}=(0,0)$ and $\bm{b}=(-0.3,0)$, for which the WCCs  $\bar{x}_j(k_y)$ anticross, as shown in \wccapp. The discussion of \cref{CMy} proves that the edge states are protected by $\mathcal{CM}_y$ symmetry, which arise from the crystalline symmetry $\mathcal{R}_{2x}$ (multiplied with the effective 2D chiral symmetry $\mathcal{CM}_z$, see \cref{tablesymm}). This is the only crystalline symmetry of the 1D system, however the bulk (and therefore the WCCs, too) possesses  two  additional symmetries: mirror reflection $\mathcal{M}_x$ and inversion $\mathcal{I}$. The anticrossing of the WCCs is, in fact, owed to inversion symmetry $\mathcal{I}$, and this feature survives even when $\bm{b} \nparallel \hat{\bm x}$,  since the magnetic field pseudovector $\bm{b}$ is invariant to inversion. If we have additionally $\bm{s}=(0,0.3)$ [while $\bm{b}=(-0.3,0)$], then only $\mathcal{M}_x$ survives in the bulk  and thus both the 1D  ribbon and the WCCs (\wccxapp) are gapped.

We remark that $\bar{x}(k_y)$ in \wccxapp~are fully gapped when $\bm{s} =(0,s_y)$, but this spin-singlet pairing only removes the anticrossing $\bar{y}(\pi)$.  The other crossing, $\bar{y}(0)$, is \emph{not} protected by any symmetry and can be annihilated  by additionally considering an infinitesimal $s$-wave pairing $s_0\ll|s_y|$ in $\Delta_{\bm k,\downarrow\uparrow}$.\\

\textbf{The nested WLO.} If one can separate two sets of Wannier bands which do not touch over the entire range $k_y \in [0,2\pi/a]$, one set can be conventionally called the "occupied" Wannier subspace and can be topologically characterized separately. The nested WLO,  introduced in Ref.~\cite{Benalcazar2017} and presented below in \cref{eq:wlo2app}, 
can be computed using the "occupied" kets $\ket{\bar{x}_n(\bm k)}$ (only one in our model), by first constructing the linear combinations
\begin{equation}
	\ket{w^n_{x,\bm k}} = \sum_{j=1}^{N_\text{occ}}  \big[\bar{x}_n(\bm k)\big]^j \ket{u^j_{\bm k}}, \qquad n \in \{1,...,N_\text{occ}/2\}.
\end{equation}

The overlap matrices  are, similarly to \cref{Oxk},
\begin{equation}
	\left[\widetilde{O}_{xy,\bm k}\right]_{nm} := \braket{w^n_{\bm k} | w^m_{\bm k -\Delta_y}} = \delta_{nm} + i \Delta \, \left[\tilde{\mathcal{A}}^{xy}_{\bm k}\right]_{nm} + \mathcal{O}(\Delta^2),
\end{equation}
where $\Delta_y = (0,\Delta)$ and the new Berry connection $\tilde{\mathcal{A}}^{xy}_{\bm k }$ has elements $[\tilde{\mathcal{A}}^{xy}_{\bm k }]_{nm} = i \bra{w^n_{x,\bm k}} \partial_{k_y} \ket{w^m_{x,\bm k}}$. The corresponding WLO follows \cref{eq:wlo1app},
\begin{equation}
	\widetilde{\mathcal{W}}_{xy,k_x} =\lim_{N\to\infty} \overline{\prod_{j}}  \widetilde{O}_{xy,\bm k + j\Delta_y} = \overline{\exp}\left(i\oint_{q_x =k_x} \mathrm{d} q_y\; \tilde{\mathcal{A}}^{xy}_{\bm q }  \right)
	\label{eq:wlo2app}
\end{equation}
and may be regarded as the WLO of the edge Hamiltonian $\mathcal{W}_{x,\bm k}$. The second-order topological invariant reads
\begin{equation}
	\phi_{xy} = \frac{1}{N}\sum_{k_x}\arg[\det\widetilde{\mathcal{W}}_{xy,k_x}] = \frac{a}{2\pi} \int_0^{2\pi/a} \!\!\!\mathrm{d}k_x\,\arg[\det\widetilde{\mathcal{W}}_{xy,k_x}]\,.
	\label{eq:phixyapp}
\end{equation}

In our model, $N_\text{occ}=2$ and $\widetilde{\mathcal{W}}_{xy,k_x}$ is, in fact, a complex number (not a matrix), such that no "$\det$" is needed and the topological invariant is encoded by a single Wannier band.  

The other topological invariant, $\phi_{yx}$, is computed in a similar fashion, by formally swapping the labels $x \leftrightarrow y$. It is worth mentioning that $\phi_{xy}$ and $\phi_{yx}$ are related to the the corner polarization~\cite{Benalcazar2017,Benalcazar2017b} (up to a factor $2\pi/e$, where $e$ is the electron charge).


\begin{figure}[t]
	\centering
	\includegraphics[width=.92\linewidth]{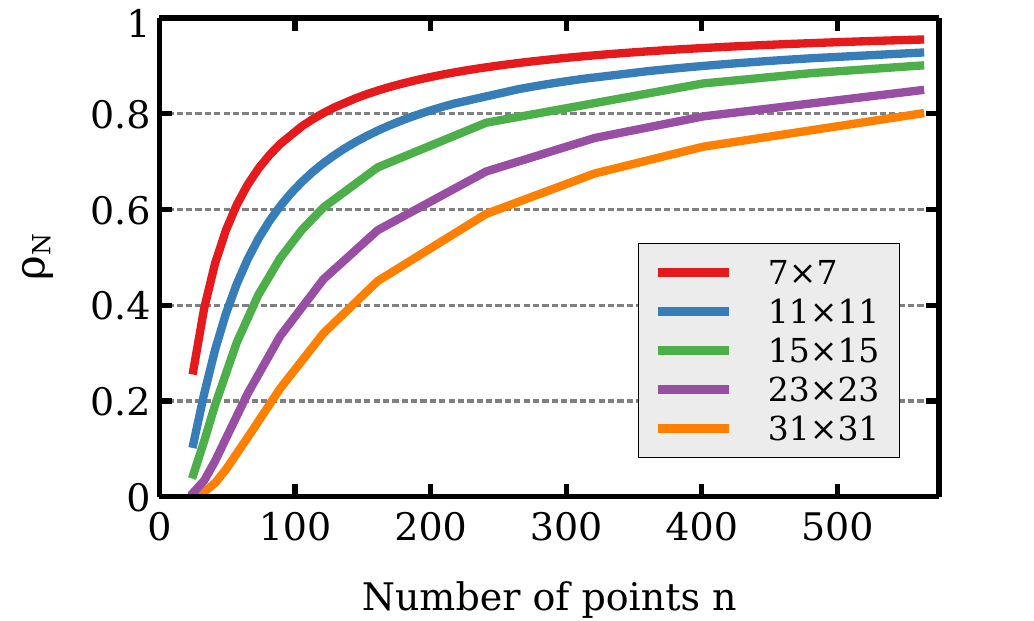}
	\caption{Non-unitarity $\rho_N$ of the WLO in $t$-space, as a function of the number of numerical steps, for various material sample sizes $N\times N$. In Eq.~\eqref{eq:rhoNnWLO} we defined $\rho_N(n)$ for either of the two corner states (it has the same behavior for both A and B).}
	\label{fig:overlaps_vs_nN}
\end{figure}

\newcommand{\rhonN}{\cref{fig:overlaps_vs_nN}}


\section{The statistical phase}

\label{app:statphase}

We computed the statistical phase as a Berry phase in $t$-space -- see \cref{eq:wlo1app} for the Brillouin zone WLO. We evaluated the corresponding WLO by making a discretization of the $t$-space in $n$ steps:
\begin{equation}
	\rho_N(n) \; \exp \left(i \Phi_\alpha\right)= \overline{\prod_{j=\overline{1,n}}} \braket{\,\psi_\alpha( t_{j-1})\,|\,\psi_\alpha( t_{j})\, }
	\,,\quad t_j = j  \frac{T}{n}.
	\label{eq:rhoNnWLO}
\end{equation}

Here $\alpha\in \{\text{A},\text{B}\}$ refers to either of the two MCSs and the factor $\rho_N(n)$ -- independent of $\alpha$, but depending on the system size $N$ and on the density of the time mesh $n$ -- has been introduced to account for the non-unitarity  of the Wilson loop product on the right-hand side (note that $N$ is finite, thermodynamic limit ${N\to\infty}$ is not assumed). The periodic boundary condition $\mathcal H^{(t=0)}=\mathcal H^{(t=T)}$ implies  
$\ket{\,\psi_\alpha( t_{0})\,} \equiv \ket{\,\psi_\alpha( t_{n})\,}$.

In \rhonN~ we plot $\rho_N(n)$ for several system sizes. Not surprisingly, it increases when the number of steps $n$ increases and it would reach unity if the braiding speed went to zero (perfect adiabaticity, $\lim_{n\to\infty}\rho_N(n)=1$). On the other hand, for a certain $n$, it decreases when the system size increases; physically, this means the speed of the braiding process should be lower for larger systems because the phase space where the MCSs live is larger.

\end{appendix}


	\bibliography{paper}
	\bibliographystyle{apsrev4-1}

\end{document}